\documentclass[showpacs,preprintnumbers,amsmath,amssymb,superscriptaddress]{revtex4}

\usepackage{graphicx}
\usepackage{dcolumn}
\usepackage{bm}

\begin{document}

\preprint{J. Phys.: Condens. Matter \textbf{20}, 245102 (2008)}

\title{Coexistence of hexatic and isotropic phases in two-dimensional Yukawa systems}

\author{Wei-Kai Qi}
\affiliation{Institute of Theoretical Physics, Lanzhou University,
Lanzhou $730000$, China}

\author{Shao-Meng Qin}
\affiliation{Institute of Theoretical Physics, Lanzhou University,
Lanzhou $730000$, China}

\author{Xiao-Ying Zhao}
\affiliation{Institute of Theoretical Physics, Lanzhou University,
Lanzhou $730000$, China}

\author{Yong Chen}
\altaffiliation{Author to whom correspondence should be addressed.
Email: ychen@lzu.edu.cn}
\affiliation{Institute of Theoretical Physics, Lanzhou University, Lanzhou $730000$, China}
\affiliation{Key Laboratory for Magnetism and Magnetic materials of the Ministry of Education, Lanzhou University, Lanzhou $730000$, China}

\date{\today}

\begin{abstract}
We performed Brownian dynamics simulations on the melting of
two-dimensional colloidal crystals in which particles interact via a
Yukawa potential. A stable hexatic phase was found in the Yukawa
systems, but we also found that the melting of Yukawa systems is a
two-stage melting, which is inconsistent with the
Kosterlitz-Thouless-Halperin-Nelson-Young (KTHNY) theory. A two-phase
coexistence region between the stable hexatic phase and the isotropic
liquid phase was found. The behavior of point defects in the
coexistence region is very complicated. The emergence of some
unstable free disclinations and grain boundaries was a characteristic
representative of the isotropic liquid phase, and a large number of
free dislocations indicated the existence of a hexatic phase. This
indicates the existence of a phase of hexatic-isotropic liquid phase
coexistence. The  big picture in the melting of a two-dimensional
Yukawa system is that first the system undergoes a transition
induced by the formation of free dislocations, then it goes through a
phase coexistence, and finally, it comes into an isotropic fluid
phase. This melting process is consistent with experiments and
simulations.
\end{abstract}


\maketitle

\section{\label{sec:level1}INTRODUCTION}

In contrast to the case of melting in three-dimensional systems, it
has by now been well established that in two-dimensional (2D)
crystals, long-range positional order does not exist due to
long-wavelength fluctuations~\cite{mer}. Despite this, in a 2D
crystal, there exists a special kind of long-range bond orientational
order. A microscopic scenario of 2D melting has been posited in the
form of the Kosterlitz-Thouless-Halperin-Nelson-Young (KTHNY)
theory~\cite{two,kt,nh,Yp}. The KTHNY theory predicts a new phase,
the so-called hexatic phase,  that exists between the solid and
liquid phases in 2D melting~\cite{Dn}.

According to the KTHNY theory, the melting of a two-dimensional
system is a two-stage transition. In the first stage we start with
the two-dimensional system in a solid phase, which has both
quasi-long-range positional order and long-range bond orientation
order; the system then undergoes a continuous transition and becomes
to a hexatic phase with short-range positional order and
quasi-long-range orientational order. In the second stage, another
continuous transition drives the hexatic phase to an isotropic liquid
phase in which both positional and bond orientational order have
short ranges.

The KTHNY theory predicts the unbinding of topological defects to
break the symmetry in the two-stage transitions. The physical driving
force behind the two-stage transitions is the dissociation of bound
defect pairs, specifically pairs of dislocation (solid$\to$hexatic)
and pairs of disclinations (hexatic$\to$liquid). Two-dimensional
systems are characterized by two different order parameters, namely,
the orientational and translational order, corresponding to the two
types of topological defects. Dissociation of the dislocation pairs
causes the translational symmetry to be broken, and dissociation of
free dislocations melts causing the orientational symmetry to be
broken.

In recent years, a large number of experiments and computer
simulations have indicated that there is, indeed, a two-stage melting
scenario in two-dimensional systems as prescribed by the KTHNY
theory. The hexatic phase has been observed in colloidal
crystals~\cite{Hexatic1, Hexatic3}, magnetic bubble
arrays~\cite{Hexatic4}, and the freestanding liquid-crystal
films~\cite{Hexatic5}. The hexatic phase not only appears in 2D
systems, but also in three-dimensional systems, such as in layered
smectic liquid crystals~\cite{Hexatic6}, dense solutions of
DNA~\cite{Hexatic7,Hexatic8} and high temperature
superconductors~\cite{Hexatic9}.

Although the KTHNY theory is currently preferred, a different
theoretical approach, evoking grain-boundary-induced melting, was a
first-order transition suggested by Chui~\cite{chui}. Using a
low-density approximation, Chui found that the grain boundaries might
be generated before the dislocations unbind when the core energy of
dislocations is sufficiently small ($E_{c}\leq 2.84\textit{k}_{B}T$);
Thus, he predicted a first-order transition. One may note that the
condensation of geometrical defects is also a first-order
transition~\cite{Gd1,Gd2}.

Several computer simulations on 2D melting have favored a first-order
phase transition or, at most, a weak first-order transition. In these
simulations, the hexatic phase was not observed. It was argued that
this transition might depend on the specific properties of systems
being studied such as their inter-particle potential. The transition
seems to be first-order in hard-core systems~\cite{ja}, but to be
second-order behavior in dipole-dipole interactions~\cite{bo}. Some
simulations for Lennard-Jones systems have discovered that the
hexatic phase is metastable~\cite{Kn}. In the case of Yukawa systems,
Naidoo and Schnitker found that the defect topology was very
complicated, and the predictions of the KTHNY theory were
violated~\cite{Ns}.

The hexatic phase has now been indeed observed in 2D colloidal
crystals~\cite{zm1,My,Ta}. Two-dimensional colloidal particles are a
good experimental system for studying the 2D melting. The advantage
of such systems is that the charged colloidal crystals in aqueous
suspensions are observable under the microscope where, due to the
particle size, the individual colloid motions can be directly
observed. At the same time, the particles are still small enough to
perform thermally driven motion and can be considered as a
statistical ensemble in thermal equilibrium. In 1987, Murray \textit{et al.} experimented on these colloidal suspensions of simple
polystyrene spheres, and the results indicated that the two-step
melting was the same as prescribed by the KTHNY theory. However, the
experimental evidence for two-dimensional colloidal suspensions
remains a hotly debated subject of controversy. Tang et al. also
observed the two-step melting process, but they indicated that the
melting was first-order, which is consistent with the picture
developed by Chui.

One commonly encounters the following essential question: does the
melting of two-dimensional screened coulomb colloidal systems (or
Yukawa systems) follow the scenario of KTHNY theory? In
Ref.~\cite{Ns}, the answer is that they found an intermediate phase,
but not a true hexatic phase, in their simulations. In the other
hand, Murray et al. observerd the hexatic phase in the screened
Coulomb system. A similar result has been obtained by Tang et al.,
but it worth noticing that although the hexatic phase was observed,
the possibility of the coexistence of the hexatic phase with the
isotropic liquid phase could not be ruled out in their
experiments~\cite{soft2}.

In this paper, we study the melting of two-dimensional charged
colloidal crystals, where we present a Brownian-dynamics simulation
on a two-dimensional Yukawa system. We focus on the existence of the
stable hexatic phase and the coexistance of an isotropic-hexatic
phase. Furthermore, we study the defect structure and compare it with
former results from experiments and simulations. In Murray and Van
Winkle's work, the two-step KTHNY melting was found by a correlation
length analysis, however, the paired dislocations in the solid phase
and the free dislocations in the hexatic phase were not found. Tang
\emph{et al.} have also obtained a similarly contradictory result. By
Voronoi constructions analysis, we find that the paired dislocations
exist in the solid phase and only a few unstable free dislocations
are found in the hexatic phase.

The organization of this paper is as follows. In Sec. II, we describe
the Brownian-dynamics simulation methods. The results are presented
and discussed in section III. Here, we calculate the orientational
and pair correlation functions, and then examine the two-phase
coexistence regions and defect topology. In the last section, we
summarize our results.

\section{\label{sec:level1}The Model}
The well-known Dejaguin-Landu-Verwey-Overbeek (DLVO) theory gives us
a good description for the effective pair interaction of the
one-component model in the colloidal systems. The DLVO potential
consists of an electrostatic repulsion and a van der Waals
attraction. Normally this is managed in a Yukawa or screened Coulomb
form, which only retains the electrostatic part~\cite{Yw0}. For a
dilute charged stabilized colloidal system in which many-body
interactions can be ignored~\cite{Yw1,Yw2,Yw3}, a pairwise Yukawa
potential is defined as
\begin{equation}
V(r) = U_{0} \frac{\sigma}{r} \exp \left( - \lambda
\frac{r-\sigma}{\sigma} \right ), \label{eq01}
\end{equation}
where $U_{0}$ is the energy and $\sigma$ is the scale length. The
screening parameter $\lambda$ describes the 'softening' of the
particles: when $\lambda$ increases from zero to infinity, the cores
of interacting particles change from very soft to extremely
hard~\cite{2d,BD1,BD2}. In our simulations, we assumed that the
Yukawa potential between the particles is very soft (the screening
parameter $\lambda = 8$).

Now we briefly describe the standard Brownian-dynamics simulation,
which is based on a finite difference integration of the irreversible
Langevin equations. The equation of motion for an individual colloid
$i$ is
\begin{equation}
\xi \dot{\textbf{r}}_{i} (t) = \textbf{F}_{i} (t) + \textbf{R}(t),
\label{eq02}
\end{equation}
where $i=1,\ldots,N$ labels the $N$ particles, $\xi$ is the friction
coeffcient, and $\xi=1$ in simulation units. $\textbf{R}(t)$ is the
Langevin random force of the solvent, and $\textbf{F}_{i}(t)$ is the
total inter-particle force on particle $i$. Here the hydrodynamic
interaction is ignored. The finite difference integration is
\begin{equation}
\textbf{r}_{i}(t+\Delta t) = \textbf{r}_{i}(t) + \textbf{F}_{i}(t)
\Delta(t) + (\Delta\textbf{r})_{R} + \textit{O}(\Delta t)^2,
\label{eq03}
\end{equation}
where $(\Delta \textbf{r})_{R}$ is a random displacement sampled from
a Gaussian distribution of zero mean and variance
\begin{equation}
\overline{(\Delta\textbf{r})^2_{R}} = 4 D_{0} \Delta t. \label{eq04}
\end{equation}
Here $D_{0}=\textit{k}_{B}T/\eta$ is the short-time diffusion
coefficient, $\textit{k}_{B}$ is the Boltzmann Constant and $T$ is
the temperature. The coefficient in Eq.~(\ref{eq04}) is $2$ in
one-dimensional systems. We used reduced units such that $U_{0}=1$,
$\sigma=1$, and $\rho=N/V=1$. In all simulations, we tuned the
reduced temperature $T^{*}=k_BT/U_0$ and the other parameters
$\sigma$, $U_{0}$, $\rho$, and $\lambda$ were fixed. We used a
periodically repeated rectangular simulation box with $N=2500$
particles and started from a triple lattice. The cutoff $r_{c}$ was
set as $4.1$. We considered only a triangular lattice since it is the
most densely packed lattice in two-dimensions and is thus favored by
nature. The scale of the simulation box is in the ratio $2:\sqrt{3}$
with the length of the $x$-axis of our simulation box $55.836$ in
order to minimize the finite-size effects. One can find more details
about this simulation in Ref.~\cite{2d}.

To characterize the translational order, we calculate the pair
correlation function. It is defined by
\begin{equation}
g(r) = \rho^{-2} \left \langle \sum_{i,j \neq i}
\delta(\textbf{r}_{i}) \delta(\textbf{r}_{j} - \textbf{r}) \right
\rangle, \label{eq05}
\end{equation}
where $\rho$ is the $2D$ particle density. The bond-orientational
function is
\begin{equation}
g_{6}(r) = \left\langle \psi_{6}^{*} (\textbf{r}') \psi_{6}
(\textbf{r}' - \textbf{r}) \right\rangle, \label{eq06}
\end{equation}
where $\psi_{6}(\textbf{r})$ is the local bond orientational order
parameter
\begin{equation}
\psi_{6}(\textbf{r}_{m}) = \frac{1}{N_{b}} \sum_{n=1}^{N_{b}} e^{6i
\theta_{mn}}. \label{eq07}
\end{equation}
Here $N_{b}$ denotes the number of the nearest neighbor of the $m$th
particle, and $\theta_{mn}$ is the angle between the particles $i$
and $j$ with an arbitrary, but fixed, reference axis. According to
the KTHNY theory, the bond-orientational function $g_{6}(r)$ will
have an algebraic decay in a hexatic phase, and an exponential decay
in the liquid phase. Before the dislocation unbinding transition
occurs, the hexatic phase is anisotropic and the bond orientational
correlation function is
\begin{equation}
g_{6}(r)\propto r^{-\eta_{6}(T)}, \label{eq08}
\end{equation}
where $ \eta_{6} = \frac{18\emph{k}_{B}T}{\pi K_{A}}$. $K_{A}$ is
called the Frank Constant which describes the coupling constant
related to distortions of the bond-angle field. The KTHNY theory
predicts that the disclination unbinding transition is also
continuous, and that it occurs when the value of the Frank Constant
falls below $72 k_{B} T/\pi$.

\section{\label{sec:level1}Results and discussions}

In order to study the phase behaviors and identify the existence of
the hexatic phase, we computed the pair distribution functions and
bond orientational correlation functions. As there are large
fluctuations near the critical point, it is difficult to obtain the
phase boundaries precisely, however, what we are interested in is the
process of melting transition and phase behavior in the middle of the
intermediate region. In our simulations, it takes a sufficient amount
of time to reach equilibrium when a two-stage continuous melting
transition occurs. Normally, the simulation reaches equilibrium after
bout $50 \tau_{B}$, and the simulation results are gathered within
the range of $10 \tau_{B}$ ($\tau_{B} = \sigma^{2} \xi / U_{0}$).

\subsection{\label{sec:level2}The translational and orientational order}

\begin{figure*}
\begin{center}
\includegraphics[width=1\textwidth]{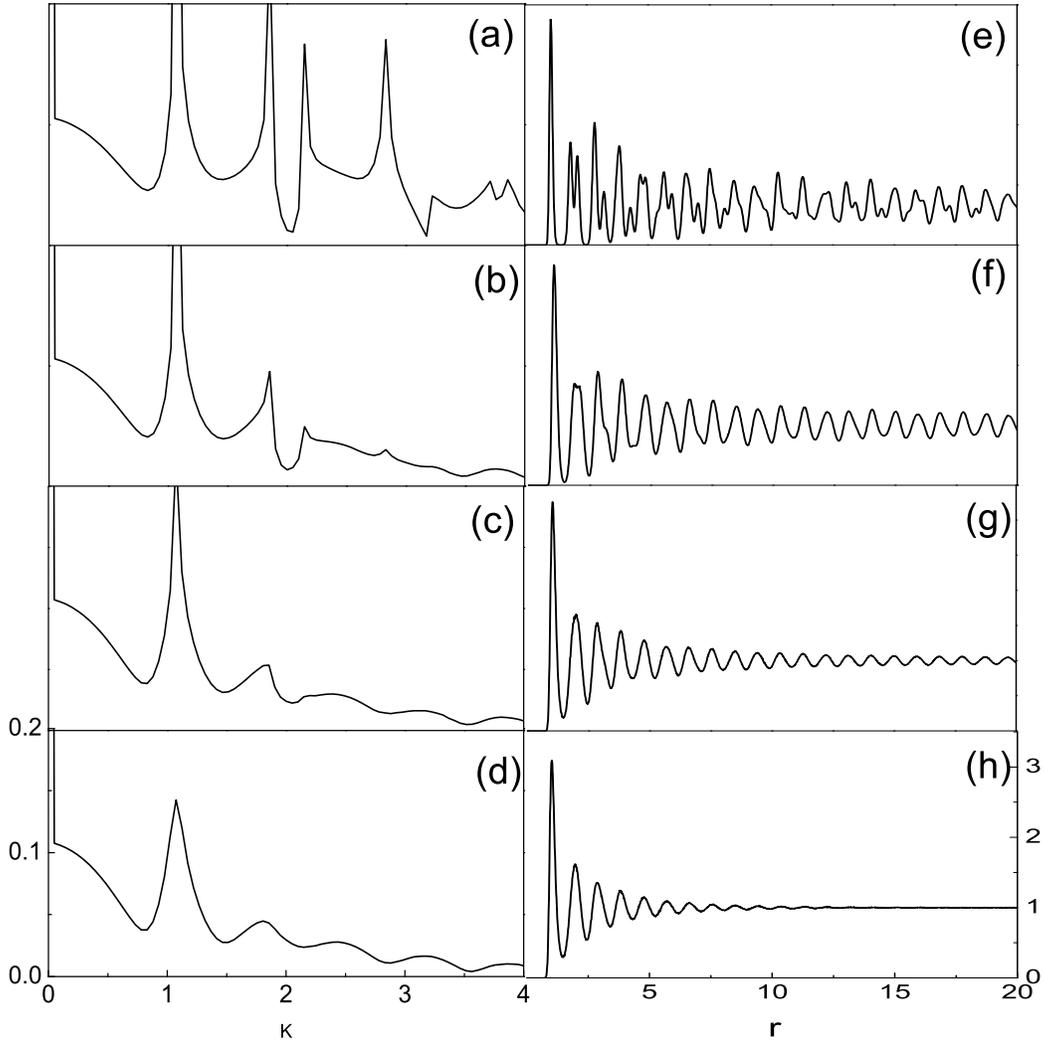}
\caption{The pair correlation function $g(r)$ and its FFT analysis in
Yukawa systems with $N=2500$ and $\rho=1.0$ for different
temperatures, (a) $T^{*}=0.200$, (b) $T^{*}=0.500$, (c)
$T^{*}=0.605$, (d) $T^{*}=0.630$.} \label{Fig:1}
\end{center}
\end{figure*}

Fig.~\ref{Fig:1}(e-h) shows the pair correlation functions $g(r)$ for
different temperatures. In this setting (the 2D crystal), there is
only an orientational symmetry; here there is no true translational
order as it can never really be "long-range". The systems are in the
solid phase for $T^{*}=0.200$ and $0.500$ and the oscillations of the
pair distribution function persist over the entire range. It was
found that the translational correlation function decays
algebraically. The behavior was different however for $T^{*}=0.605$
and $0.630$, and the oscillations died out quickly, which indicating
that short-range translational order does exist in the system.

The fast Fourier transforms (FFT) of the pair distribution functions
are also presented in Fig.~\ref{Fig:1}(a-d). At low temperature
(solid phase), the first peak that indicates the periodic structure
of the system is very sharp. The second and third peaks also are
obvious (see Fig.~\ref{Fig:1}(a, b)). As we increase the temperature,
the third peak becomes unnoticeable. For $T^{*}=0.605$, the second
peak is not obvious, but the first peak was lower than  it was for
the systems in the solid phase. At the temperature $T^{*}=0.630$, the
system is in the liquid phase. Here, none of the peaks were obvious
and the first peak was much lower than it was for the system in the
solid phase (see Fig.~\ref{Fig:1}(c, d)). With the increasing of
temperature, it show that the first peak moves considerably in
Fig.~\ref{Fig:1}(a-d). Due to crystal lattice is periodic, the first peak is
very sharp in solid phase. When the system melting into liquid phase,
the crystal lattice was broken. So it indicates that the
quis-long-range position order in solid phase became into short range
in liquid phase.
\begin{figure}
\includegraphics[width=0.5\textwidth]{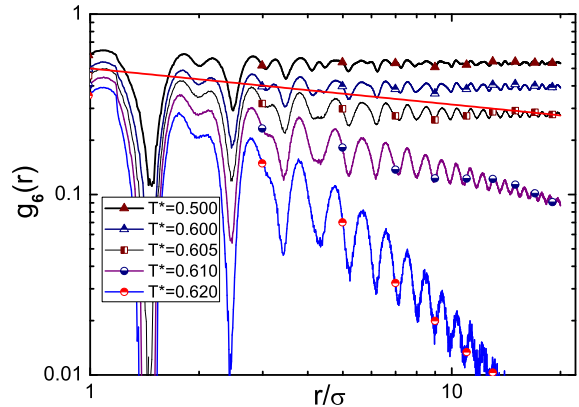}
\caption{(Color online) Orientational correlation function $g_{6}(r)$
as a function of the reduced temperature $T^{*}$ in a log-log plot.
The curves decays algebraically, which implies the existence of the
hexatic phase. The straight line, with slope $1/4$, is a guide for
the eyes.} \label{Fig:2}
\end{figure}

In Fig.~\ref{Fig:2}, we plotted the simulation results of the bond
orientational functions $g_{6}(r)$ for different reduced
temperatures. When $g_{6}(r)$ does not decay, it means that the
system is in the solid phase with a long-range bond-orientational
order. When the reduced temperature rises to near the disclination
unbinding regions, $g_{6}(r)$ decays algebraically with an exponent
near $1/4$, which implies the existence of the hexatic phase as
predicted by the KTHNY theory. As in our simulations, it is  easy to
see the hexatic phase where the bond-orientational functions decay
algebraically with $\eta(T^{*})$ near to $1/4$ (see $T^* = 0.605$ in
Fig.~\ref{Fig:2}). With further increases in temperature, the system
becomes a disordered liquid and $g_{6}(r)$ decays exponentially. It
should be noted that $g_{6}(r)$ decays algebraically with an exponent
$0.54$ at $T=0.61$, which is faster than $1/4$ from the KTHNY theory.

\begin{figure}
\includegraphics[width=0.5\textwidth]{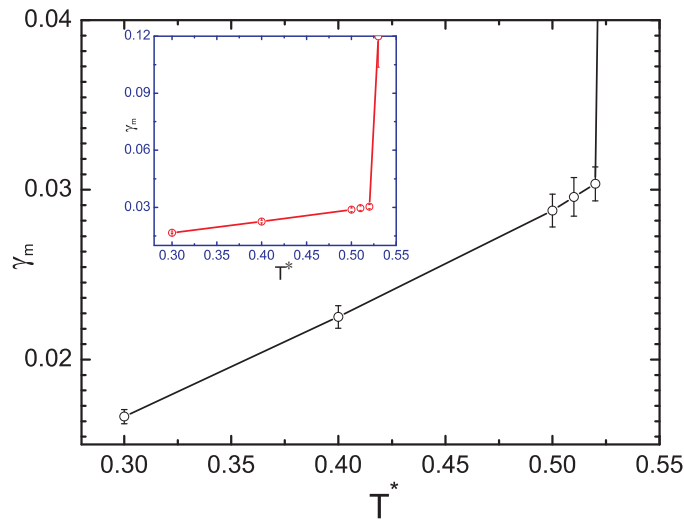}
\caption{The Lindemann parameter $\gamma_{m}$. The lines drawn are
only guides for the eyes.} \label{Fig:3}
\end{figure}

To locate the solid-to-hexatic phase transition temperature, we use
the 2D Lindemann melting criterion introduced by Bedanov and
Gadiyak\cite{lindemann}. As the mean square displacement $\langle
u^{2}\rangle$ diverges in a two-dimensional crystal, they suggested
the Lindemann parameter
\begin{equation}
\gamma_{m}=\langle(u_{j}-u_{j+1})^{2}\rangle/a^{2}
\end{equation}
where the indices $j$ and $j+1$ refer to neighboring particles. At
the melting point, these authors found a critical value
$\gamma_{m}^{c}=0.033$. Here we measure the Lindemann parameter at
different temperatures. At the melting point $T_{m}^{*}$, the
Lindemann parameter grows sharply, indicating a vanishing of the
positional symmetry. In our simulations, a sharp growth of
$\gamma_{m}$ is observed at $T_{m}^{*}=0.520$ (see Fig.~\ref{Fig:3}).

\subsection{\label{sec:level2}The coexistence of hexatic and isotropic phases}

Indeed, there exists an algebraic decay of the bond-orientational
correlation function $g_{6}(r)$. However, we observed that $\eta_{6}$
is larger than $1/4$ (see Fig.~\ref{Fig:2}), which is not consistent
with the prediction of KTHNY theory. In light of this, it has been
conjectured that a coexistence of hexatic and isotropic phases may
appear; this was pointed out by H. H. von Gr\"unberg et
al~\cite{soft2}. The coexistence region is important evidence of a
first order transition~\cite{2d}, however, in order to identify this
coexistence we need to work carefully.

\begin{figure}
\includegraphics[width=0.5\textwidth]{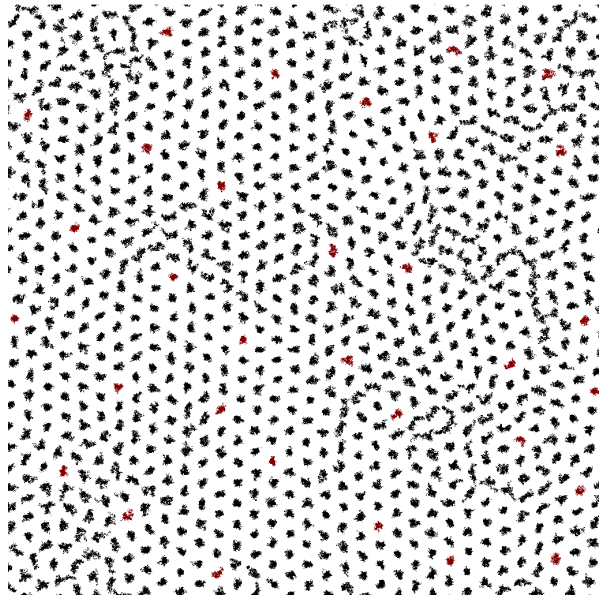}
\caption{Particle trajectory at $T^{*}=0.609$. This shows an apparent
two-phase equilibrium in two-dimension Yukawa systems. This
trajectory is plotted during $6 \tau_{B}$.} \label{Figtr}
\end{figure}

Fig.~\ref{Figtr} plots the trajectory of the particles at
$T^{*}=0.609$ for a section of the simulation box during $6\tau_{B}$.
The existence of solid-like and liquid-like patches in
Fig.~\ref{Figtr} has been interpreted as evidence for phase
coexistence, however, a similar path also appears in the purely
hexatic phase. Given this, we have to present a more reliable and
measurable method to identify whether the systems is the pure hexatic
phase or in a two-phase coexistence.

In order to confirm the phase coexistence, we used the recipe
suggested by Strandburg \textit{et al}~\cite{ch6}. The angular
susceptibility $\chi_{6}$ for different length scales is defined by
\begin{equation}
\chi_6=\bigg<\bigg|\frac{1}{N}\sum_{l}\frac{1}{N_l}\sum_{n}e^{6i\theta_{mn}}\bigg|^2\bigg>
\end{equation}
where the sum on $l$ is over all particles, the sum on $n$ is over
the nearest neighbors, $n_l$ is the number of nearest neighbors of
particle $l$, and $N$ is the number of particles in the system.In the
solid phase, the angular susceptibility $\chi_{6}$ is large due to
the long-range order. On the other hand, $\chi_{6}$ is small in the
fluid phase. If the system exhibits a two-phase coexistence, one
might expect that the distributions of $\chi_{6}$ for sufficiently
small length scales could be modeled by a combination of solid and
fluid distributions. However, in the case of a homogeneous hexatic
phase, varying the size of the subsystems should not lead to any
qualitative changes in the distribution of $\chi_{6}$ (as shown in
Ref.~\cite{bo} for systems with dipole-dipole interactions).

We calculate $\chi_{6}$ in our simulations for many different length
scales by dividing the system of $2500$ particles into subsystems
containing an average of $128$, $64$, $16$, and $4$ particles. These
subsystems are also periodically repeated as rectangular simulation
boxes. $\chi_{6}$ is calculated every $100$ passes. If the system is
homogeneous, varying the size of the subsystems should not lead to
any qualitative changes in the distribution of $\chi_{6}$.
Contrarily, for a two-phase region, the probability distribution of
$\chi_{6}$ at a sufficiently small length scale could be modeled by a
curve with two peaks, reflecting a combination of two-phase
distributions.

\begin{figure*}
\begin{center}
\includegraphics[width=0.9\textwidth]{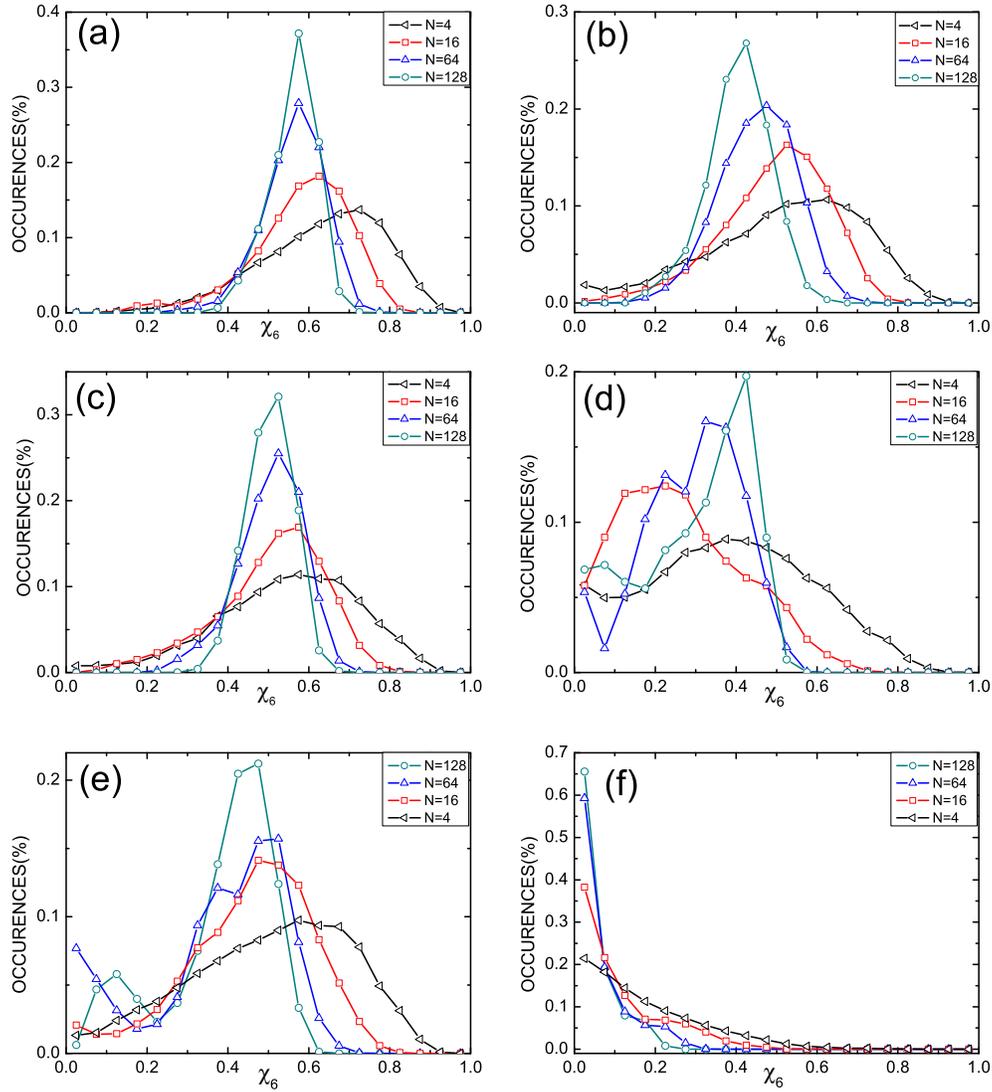}
\caption{The probability distributions of $\chi_{6}$ for different
temperatures: (a) $T^{*}=0.500$, (b) $T^{*}=0.600$, (c)
$T^{*}=0.602$, (d) $T^{*}=0.605$, (e) $T^{*}=0.609$, and (f)
$T^{*}=0.630$. The symbols in the plots indicate the number of
particles in different subsystems.} \label{Figk}
\end{center}
\end{figure*}

Fig.~\ref{Figk} shows the distributions of $\chi_{6}$ for subsystems
of $128$, $64$, $16$, and $4$ particles with different temperatures.
In the region of low temperature or crystal phase, there is no
qualitative change (see Fig.~{\ref{Figk}}(a)). The probability
distributions of $\chi_{6}$ always remain singly peaked as the size
of the subsystems is varied. This indicates that the system is in the
homogeneous solid phase for low temperatures. In Fig.~\ref{Figk}(b)
and (c), we presented the probability distribution of $\chi_{6}$ at
$T^{*}=0.600$ and $T^{*}=0.602$ where a power law decayed of
$g_{6}(r)$ was found. By varying the size of the subsystems, the
probability distribution of $\chi_{6}$ has a single peak, which means
that there is only a homogeneous phase and the two-phase coexistence
can be ruled out. One can conclude that there is a stable hexatic
phase at $T^{*}=0.600$ and $0.602$.

In Fig.~\ref{Figk}(d) and (e), the probability distributions of
$\chi_{6}$ at $T^{*}=0.605$ and $0.609$ could be modeled by a curve
with two peaks for $n=64$ and $128$. This clearly demonstrates the
two-phase coexistence since the peaks reflect a combination of solid
and fluid distributions. It should be noted that the $g_{6}(r)$
decays algebraically at both temperatures.

Fig.~\ref{Figk}(f) depicts the probability distribution of $\chi_{6}$
at $T^{*}=0.630$. In the fluid phase, the distributions are peaked
near zero. Varying the size of the subsystems, the distribution of
$\chi_{6}$ cannot lead to any qualitative changes so we can conclude
that the 2D Yukawa systems melt to a pure liquid phase.

\subsection{\label{sec:level2}Topological defects}

\begin{figure}
\includegraphics[width=0.4\textwidth]{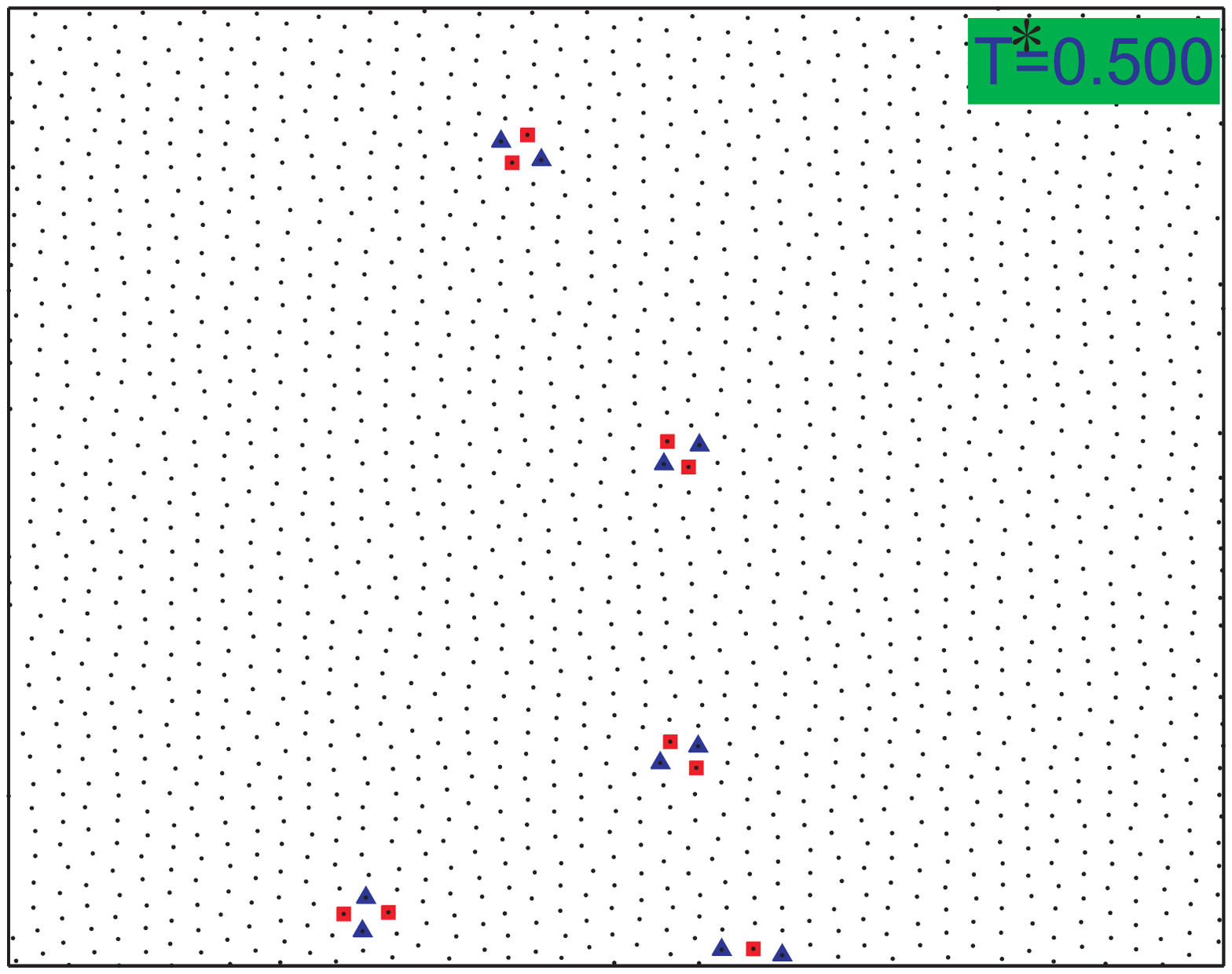}\\
\includegraphics[width=0.4\textwidth]{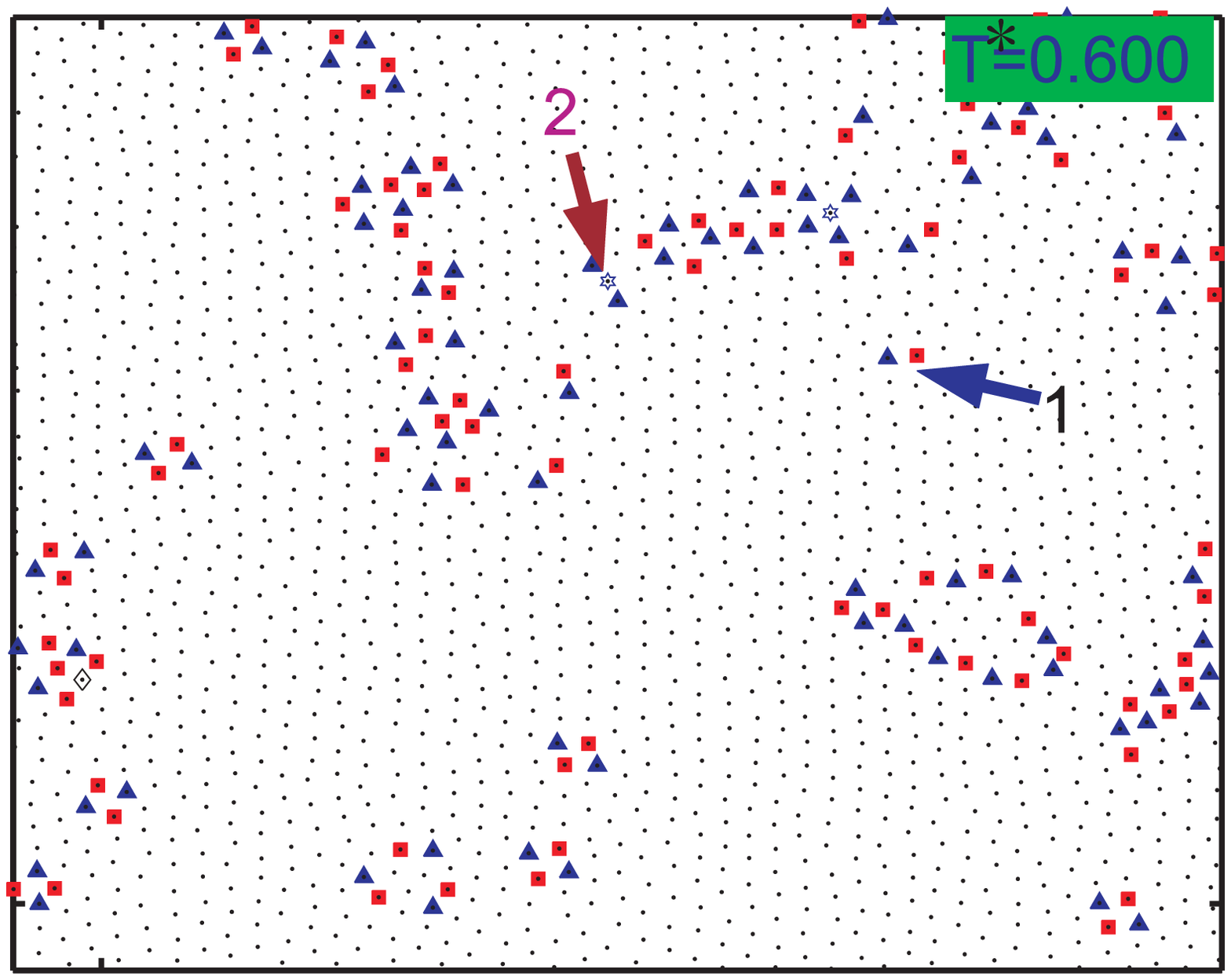}\\
\includegraphics[width=0.4\textwidth]{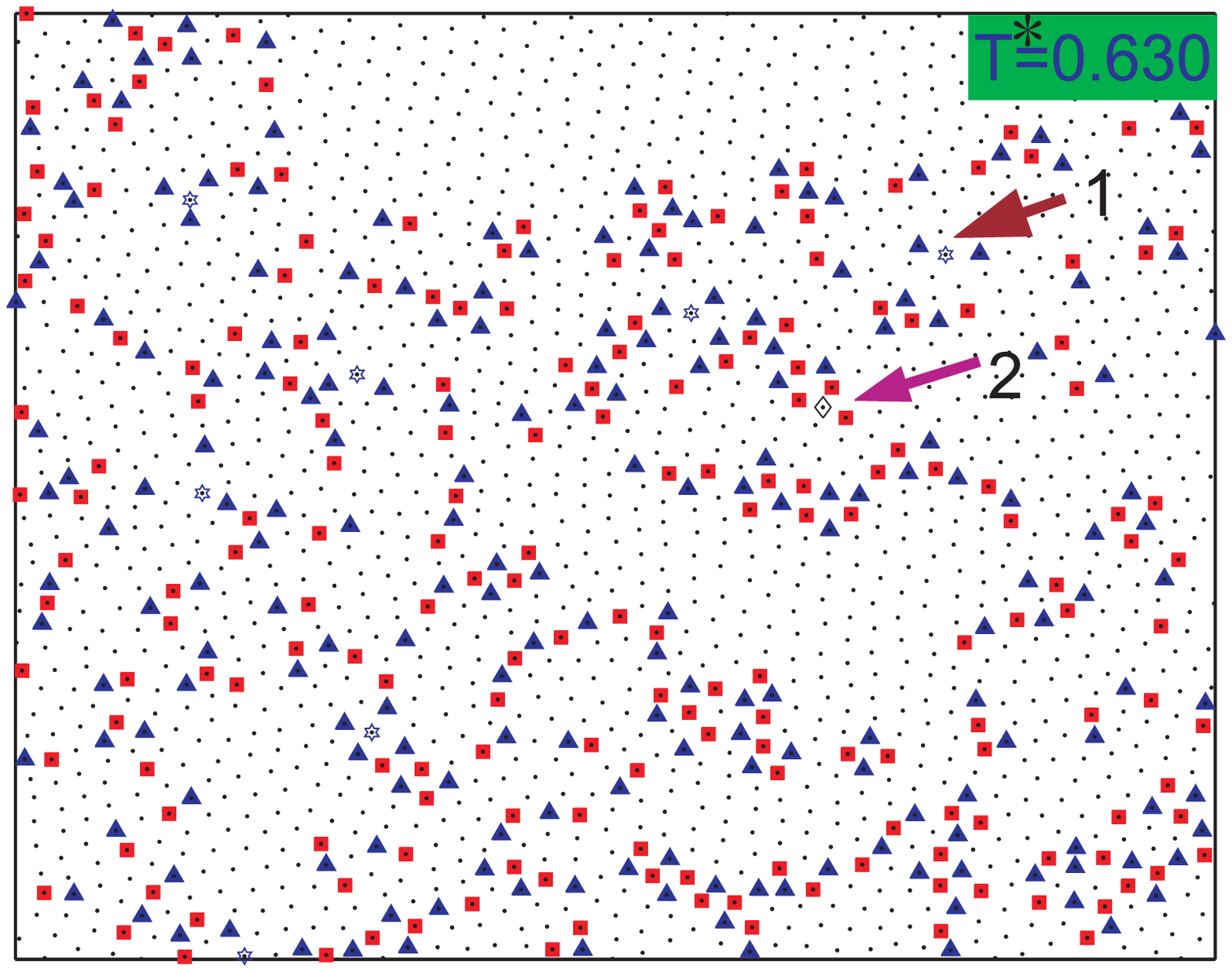}\\
\caption{Distribution of topological defects at $T^{*}=0.50$ (top),
$0.600$ (middle), and $0.630$ (down). The $\vartriangle$ denotes a
disclination of unit positive strength (five nearest neighbors) and
the $\square$ denotes a disclination of unit negative strength (seven
nearest neighbors).} \label{Figtr2}
\end{figure}

In the KTHNY theory, the two-dimensional melting is caused by the
unbinding of topological defect pairs~\cite{kt,nh,Yp}. First,
dislocation pairs unbind and then, the disclination pairs unbind.
Since dislocations are associated with an additional half row of
atoms, they can be quite effective at breaking up translational
order. However, dislocations are less disruptive of orientational
correlations. And the disclination breaks the orientational order.
Thus it is important to determine whether these defects are indeed
important in the melting process. Murray et al. observed that islands
of sixfold coordinated particle are surrounded by a network of grain
boundaries of fourfold and sevenfold coordinated particles in the
fluid phase~\cite{Hexatic9}. In the hexatic phase, they failed to find
free dislocations, but these grain boundaries did not completely
disappear. The neighboring grains began to orient with respect to
each other. In solid phase, they also failed to find paired
dislocations.

Normally, the defect structure can be pictured by using Voronoi
polygons where we identify disclinations as particles with five or
seven nearest neighbors. A Voronoi polygon is defined as the boundary
of a region enclosing a particle, which is closer to every point of
the region than to any others~\cite{V1,V2}. A disclination is located
at a particle with five or seven vertices in its Voronoi polygon. In
a perfect triangular colloidal crystal, all particles are sixfold.
Disclinations are described as particles having five and seven
neighbors. A disclination with positive unit strength is located at a
particle with five near neighbors and that with negative unit
strength is located near seven neighbors. A dislocation may be viewed
as a tightly bound pair of disclinations. In the solid phase,
dislocations are bound very strongly to a potential that increases as
the square of the separation.

We studied the topological configurations by using Voronoi cell
pictures to identify the position of the defect. We did this by
showing the positions of disclinations for a single configuration
after a long run. For a single configuration in a solid phase at
$T^{*}=0.500$ (see Fig.~\ref{Figtr2}(a)), nearly all of the
dislocations occur in pairs. Fig.~\ref{Figtr2}(b) shows the defect
structure in an intermediate region between the isotropic liquid and
the solid at $T^{*}=0.600$. It was found that there exist free
dislocations (see arrow $1$), which supports the KTHNY theory that
melting is the unbinding of dislocation pairs.

In Fig.~\ref{Figtr2}(c), the system is clearly in an isotropic liquid
phase. There are a large number of defects and free disclinations,
and the defect structure is very complicated. Clustering of
dislocations is observed~\cite{Ns}: there exists a pair of fivefold
coordination (5-coordination) bindings with an 8-coordination (shown
in Fig.~\ref{Figtr2}(c) by arrow $1$) or a pair of 7-coordination
bindings with a 4-coordination (in Fig.~\ref{Figtr2}(c) shown by
arrow $2$). These topological defects are unstable, however, and
vanish very quickly.

\begin{figure}
\includegraphics[width=0.4\textwidth]{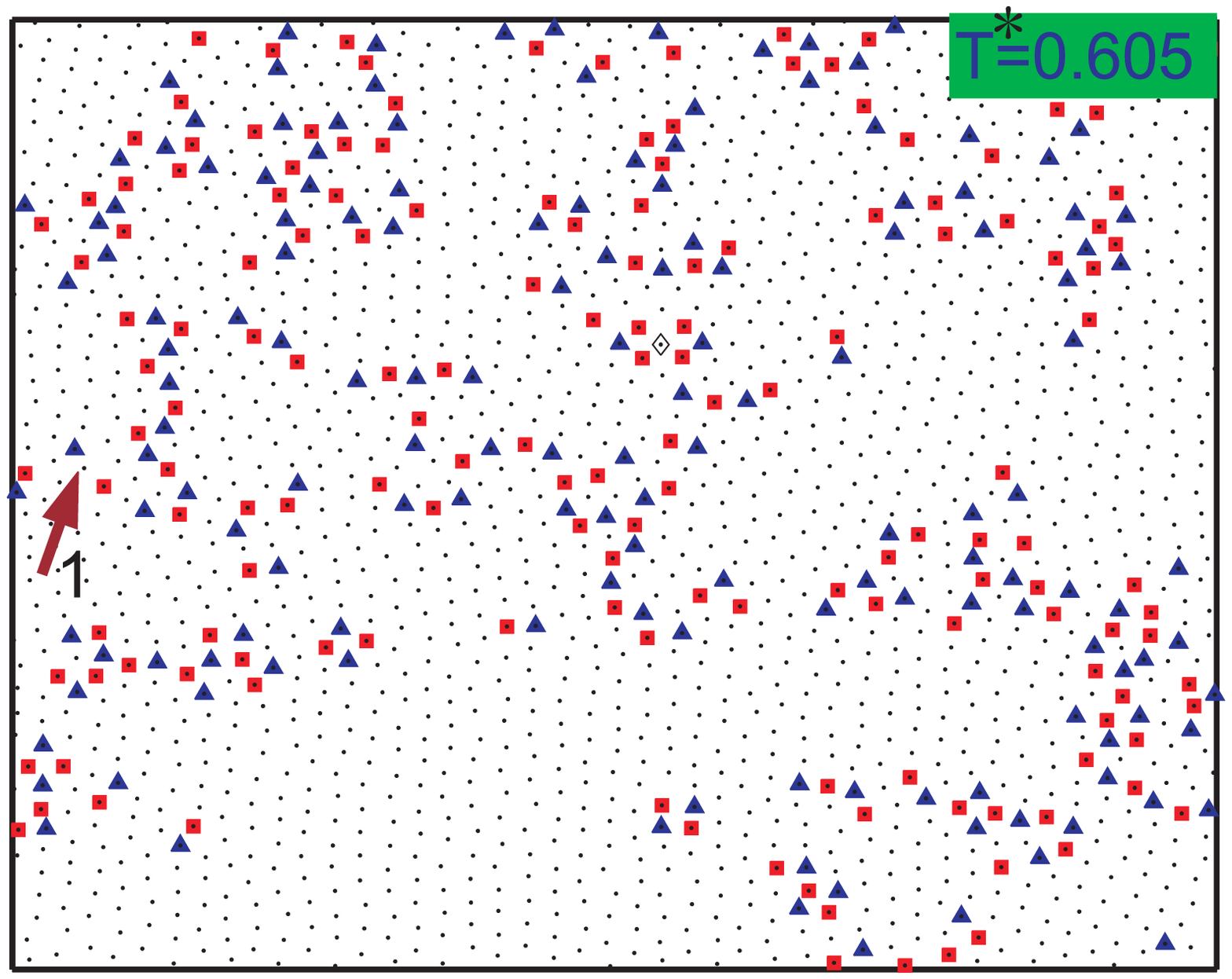}
\caption{Distribution of topological defects at $T^{*}=0.605$, where
the system is in a two-phase coexistence. A small amount of the free
dislocations appeared. $\vartriangle$ denotes a disclination of unit
positive strength (five nearest neighbors), and $\square$ denotes a
disclination of unit negative strength (seven nearest neighbors).}
\label{Figtr3}
\end{figure}

The distributions of topological defects in the two-phase coexistence
region are shown in Fig.~\ref{Figtr3}. There is a large number of
free dislocations, and only a few free disclinations. Notice that the
translational order is broken by the emergence of free dislocations.
Since the number of disclinations is not enough to break up the bond
orientational order, the bond orientational function still decays
algebraically at the two-phase coexistence region. We observe that
defects are likely a characteristic of grain boundaries, that is, the
clusters consisting of dislocations and dislocations pair up, or
small dislocations form a loop, such defects were also shown in
Tang's work. The appearance of grain boundaries leads to the
first-order transition suggested by Chui~\cite{chui}.

\begin{figure}[t]
\includegraphics[width=0.5\textwidth]{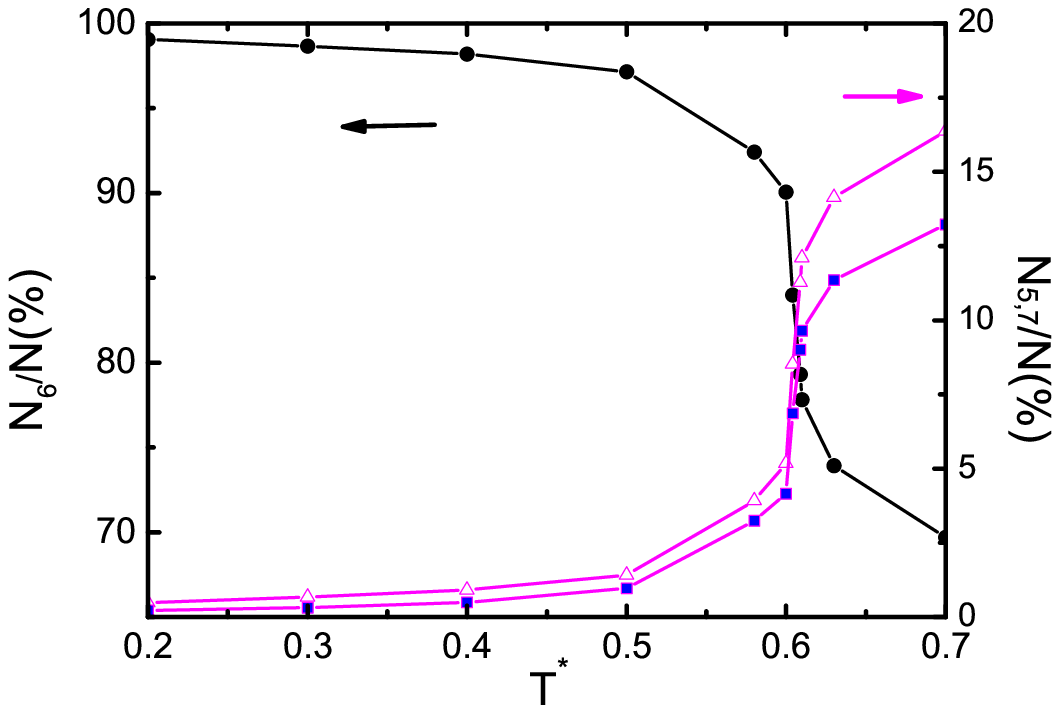}
\caption{The fractions of $6$-coordinated, $5$-coordinated, and
$7$-coordinated particles. At low temperature, the number of defects
is very small. When the temperature above $0.500$, $N_{6}/N$
decreases rapidly.} \label{Figd2}
\end{figure}

Fig.~\ref{Figd2} plots the fractions of $6$-coordinated,
$5$-coordinated, and $7$-coordinated particles. At low temperature,
all particles are nearly 6-coordinated, and the number of defects is
very small. When the temperature reaches $0.500$, $N_{6}/N$ shows
rapidly decreasing behavior. At $T^{*}=6.05$, almost $20\%$ of
particles are attached to the defects, which is consist with the
results in Ref.~\cite{Ta}. As the defect fraction rises above $30\%$,
the system melts into a liquid phase. At the liquid phase, the number
of $5$-coordinated particles is much more than the number of
$7$-coordinated particles due to the emergence of the $8$-coordinated
particles and the effect of boundaries.

We have observed the mechanism of defects in the two-dimensional
Yukawa system~\cite{mv}. At low temperature, the paired dislocation
is formed or annihilated. The formation of a binding dislocation pair
can be viewed as the simultaneous formation of two sevenfold
coordinated particles and two fivefold coordinated particles from
four sixfold coordinated particles. When the temperature rises to the
hexatic phase, the dislocation pair dissociates. In the solid phase,
the unbinding of dislocations is unstable, and these unbound
dislocations will quickly bind. This is in contrast to the stable
free dislocation is found in the hexatic phase. Grain boundaries and
unstable disclinations appears throughout the region of two-phase
coexistence. Given this, one may conjecture that the isotropic liquid
phase can be characterized by the existence of stable defect
clustering.

\section{\label{sec:level1}Conclusions}

In this paper, we performed Brownian dynamics simulations to study
the melting of 2D colloidal crystals with Yukawa interactions, and
two-stage melting is found. The hexatic phase in melting of 2D
charged colloidal crystals was indeed observed~\cite{My}. Moreover,
the hexatic-liquid phase coexistence was observed as well. Such
coexistence was also confirmed by Tang \textit{et al.}~\cite{Ta}.

We calculated the pair correlation function and bond-orientational
correlation functions. At low temperature, due to the
quasi-long-range positional order in 2D systems, the oscillations of
the pair correlation function persist over the entire range. On the
other hand, the oscillations of the pair correlation function died
quickly at higher temperature, and it was shown that the positional
order becomes shot-range. By using the 2D Lindemann melting
criterion, we found that the melting temperature is
$0.530$($\pm0.01$). An algebraic decay with $\eta$ near $1/4$ of the
bond orientational correlation function was observed at the
temperature $0.605$($\pm0.01$). By ruling out the coexistence, we
verified that this is a pure phase at the temperatures between
$0.530$ and $0.605$. As the bond orientational correlation function
decays algebraically, we concluded that the pure phase is a stable
hexatic phase with the quasi-long-range bond-orientational order.

We found that the quasi-long-range bond-orientational order still
exists in the coexistence region in finding the algebraic decays of
the bond orientational functions. The emergence of unstable free
disclinations and grain boundaries is a characteristic representative
of an isotropic liquid phase, and a large number of free dislocations
 is a characteristic representative of a hexatic phase. This indicates that
 there was indeed a coexistence of hexatic-isotropic liquid phases. In a word,
the melting of two-dimensional Yukawa systems is a two-stage melting.
Firstly, the system first undergoes a transition induced by the
formation of free dislocations, the system then goes through a phase
coexistence, and finally moves into an isotropic fluid phase.

\begin{acknowledgments}
X.Y.Z. acknowledges financial support of the National Talent Training
Fund in Basic Research. Y.C. was supported by the SRF for ROCS, SEM,
and by the Interdisciplinary Innovation Research Fund for Young
Scholars, Lanzhou University.
\end{acknowledgments}

\end{document}